\documentclass{an}
\usepackage{graphicx}
\usepackage{times} 
\usepackage{fancyhdr}
\begin{document}

\title{Faraday rotation variations along radio jets: the magnetic field in
  galaxy and group halos}

\author{R.A. Laing\inst{1}, J.R. Canvin\inst{2}, W.D.Cotton\inst{3},
  A.H. Bridle\inst{3} \and P. Parma\inst{4}}
\institute{
European Southern
Observatory, Karl-Schwarzschild-Stra\ss e 2, 85748 Garching-bei-M\"{u}nchen,
Germany
\and
School of Physics, University of Sydney, A28, Sydney,
NSW 2006, Australia
\and
National Radio Astronomy Observatory, 520 Edgemont Road,
Charlottesville, VA 22903-2475, U.S.A.
\and
INAF -- Istituto di Radioastronomia, via Gobetti 101,
40129 Bologna, Italy
}

\date{Received; accepted; published online}

\abstract{Our modelling of FR\,I radio jets as decelerating, relativistic flows
allows us to derive their orientations accurately. We present images of Faraday
rotation for two of these these objects (3C\,31 and NGC\,315) and show that the
fluctuations of rotation measure (RM) are larger in the fainter (receding) jets,
as expected if the rotation occurs in the hot galaxy/group halos.  The gas
density is much lower in NGC\,315 and the RM fluctuations are only just
detectable.\keywords{galaxies: jets -- radio continuum:galaxies -- magnetic fields
-- polarization -- MHD}}

\correspondence{rlaing@eso.org}

\maketitle

\section{Introduction}

In our models of FR\,I radio jets as relativistic flows (Laing, Canvin \& Bridle
2006), the observed differences in brightness and linear polarization between
the jets close to the nucleus are produced by relativistic aberration and
Doppler beaming and we can determine the inclination, $\theta$. Given that we
know the geometry and the external density profile (from X-ray observations),
imaging of Faraday rotation measure (RM) can determine the distribution of
magnetic-field irregularities in the surrounding hot plasma.  In this paper, we
summarize our RM imaging for two sources: 3C\,31 (Laing \& Bridle 2002) and
NGC\,315 (Canvin et al. 2005).

\section{3C\,31}

\begin{figure}
\resizebox{\hsize}{!}  {\includegraphics[]{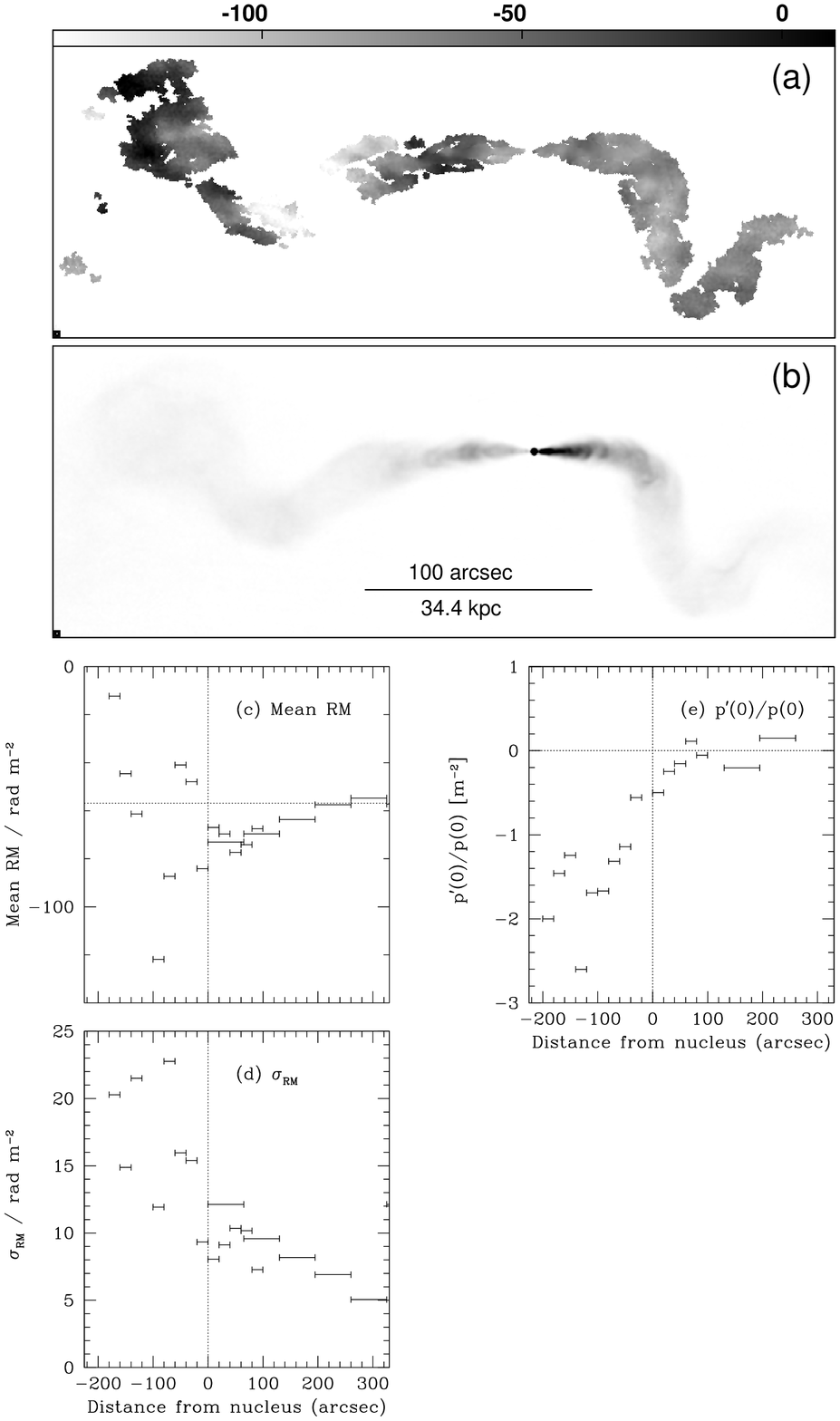}}
\caption{RM and depolarization for 3C\,31. All diagrams have the approaching
  (brighter) jet on the right.(a) RM image at a resolution of 1.5\,arcsec, made
  using the Pacerman algorithm of Dolag et al. (2005). (b) $I$ image of the same
  area.  (c) Profile of mean RM, determined in boxes along the jet axis. (d) The
  rms RM, calculated over the same areas as in panel (c). (e) Profile of
  $p^\prime(0)/p(0)$, as described in the text.}
\label{fig:3c31}
\end{figure}

3C\,31 is an FR\,I radio galaxy at a redshift of 0.0169 (0.344 kpc/arcsec for
H$_0$ = 70 km\,s$^{-1}$\, Mpc$^{-1}$). Our models of the inner jets give an
inclination of $\theta = 52^\circ$. Our RM image, derived from 6-frequency
observations between 1.365 and 8.4\,GHz, is shown in Fig.~\ref{fig:3c31}(a),
with an $I$ image at the same resolution for comparison in
Fig.~\ref{fig:3c31}(b).  The ${\bf E}$-vector position angle is accurately
proportional to $\lambda^2$ everywhere, indicating foreground rotation. There is
structure in the RM image on a range of spatial scales from 5 to $>$50\,arcsec
and the RM fluctuations are higher by a factor of 2 -- 3 on the counter-jet side
(Fig.~\ref{fig:3c31}c and d).  There is a small amount of depolarization on the
counter-jet side. We estimate this by making a first-order linear approximation
to the variation of degree of polarization, $p$, with $\lambda^2$:
$p(\lambda^2)/p(0) \approx 1 + [p^\prime(0)/p(0)]\lambda^2$
(Fig.~\ref{fig:3c31}e).  We model the thermal X-ray emission as the sum of two
components: one associated with the galaxy (core radius $r_c =$ 3.6\,arcsec),
the second with the surrounding group, with $r_c =$ 154\,arcsec (Hardcastle et
al. 2002). The Faraday rotation variations are therefore plausibly associated
with the group component. Applying the model described by Laing et al.\ (2006),
we can estimate the central magnetic field strength: $B_0$/nT $\approx$
0.9(l/kpc)$^{-1/2}$, where $l$ is the correlation length of the field.  

\section{NGC\,315}

\begin{figure}
\resizebox{\hsize}{!}  {\includegraphics[]{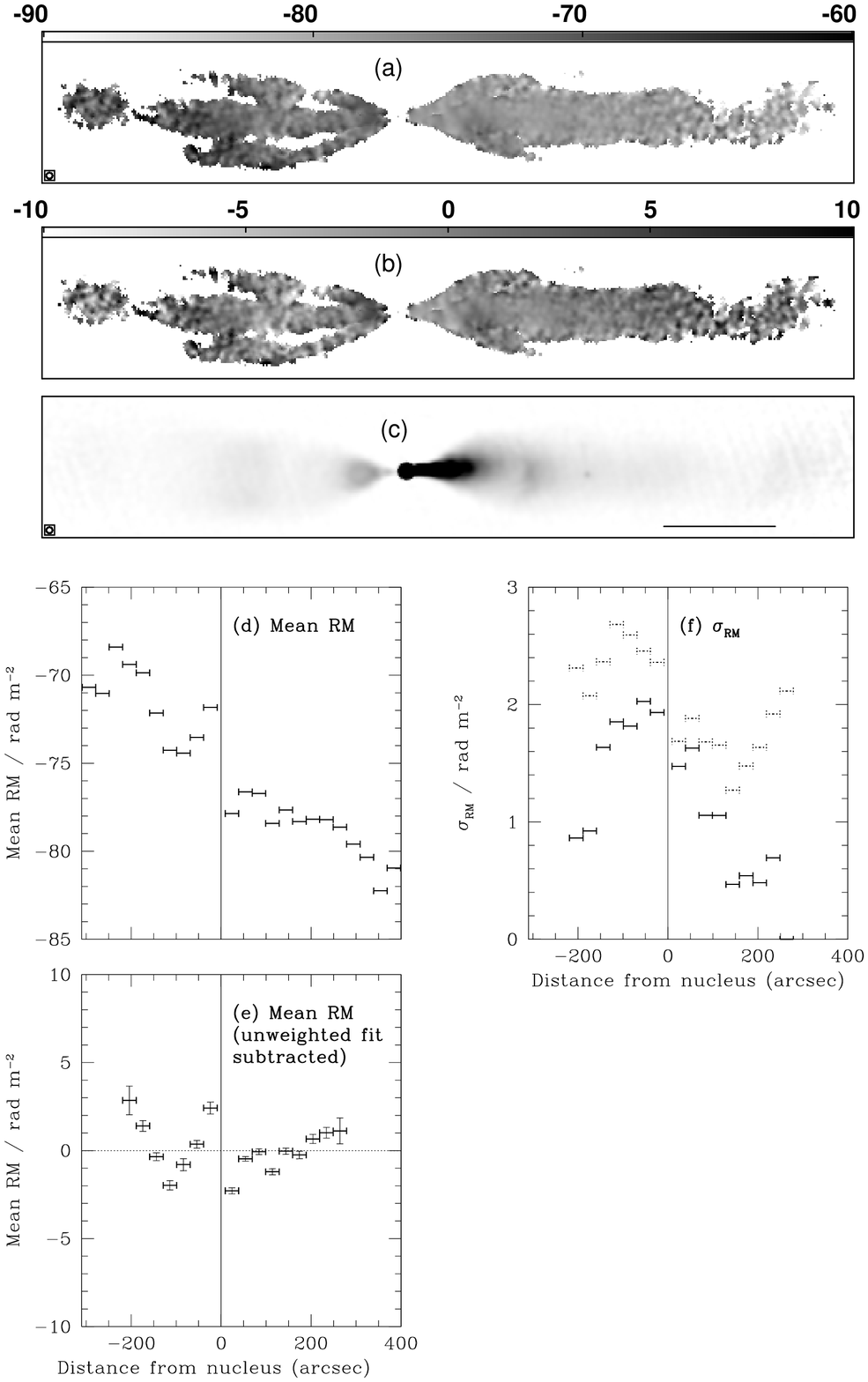}}
\caption{RM data for NGC\,315. All diagrams have the approaching jet on the
  right. (a) RM image at a resolution of 5.5\,arcsec. (b) As (a), but with a
  linear variation subtracted. (c) $I$ image covering the same area (a scale of
  100 arcsec or 34.4 kpc is indicated by the horizontal line). (d) Profile of RM
  along the jet axis. (e) As (d), but with a linear variation subtracted and
  with low s/n data omitted. (f) The rms RM for the same data points as in panel
  (e). The dotted lines show the raw values; the full lines are the values after
  a first-order correction for fitting error. Full details are given by Laing et
  al.\ (2006).}
\label{fig:ngc315}
\end{figure}

The giant FR\,I radio galaxy NGC\,315 is at a redshift of 0.01648 (0.335
kpc/arcsec). We infer an angle to the line of sight of 38$^\circ$ (Canvin et
al. 2005).  Our RM images, derived from 5-frequency observations in the range
1.365 - 5\,GHz, are shown in Fig.~\ref{fig:ngc315}, together with profiles along
the jet axis.  The mean RM and a linear gradient along the jet are almost
certainly Galactic. After removing these components, we detect RM fluctuations
on scales of 10 -- 100\,arcsec, but the typical amplitudes are only
$\approx$2\,rad\,m$^{-2}$ (Fig.~\ref{fig:ngc315}e and f). The thermal X-ray
emission from NGC\,315 has a very small core radius, $r_c = 1.55$\,arcsec
(Worrall, Birkinshaw \& Hardcastle 2003). No X-ray emission has yet been
detected from the galaxy group associated with NGC\,315 (Miller et al. 2002),
but the small amplitude of the RM fluctuations and the fact that they are larger
for the receding jet are both consistent with an origin in a tenuous group gas
component with a scale size $\approx$200\,arcsec. As we do not know the density
of this component, we can only constrain the product $(n_0/{\rm m}^{-3})^2
(B_0/{\rm nT})^2 (l/{\rm kpc}) \approx 700$, where $n_0$ is the central density
of the group component and $B_0$ and $l$ are again the central field strength
and correlation length, respectively (Laing et al.\ 2006).

\acknowledgements 

The National Radio Astronomy Observatory is a facility of the National Science
Foundation operated under cooperative agreement by Associated Universities,
Inc. We thank Corina Vogt, Klaus Dolag and Greg Taylor for the use of their RM
software.

\end{document}